\documentclass{ws-ijgmmp}
\usepackage[utf8]{inputenc}
\usepackage[english]{babel}

\begin{document}

\markboth{Artyom V. Astashenok, Karim Mosani, Sergey D. Odintsov,
Gauranga C. Samanta} {Gravitational Collapse in General Relativity
and in $R^2$-gravity: A Comparative Study}

\title{Gravitational Collapse in General Relativity and in $R^2$-gravity: A Comparative Study}

\author{Artyom V. Astashenok$^{1+}$, Karim Mosani$^{2}$, Sergey D. Odintsov$^{3,4}$, Gauranga C. Samanta$^{2}$}

\address{$^{1}$Institute of Physics, Mathematics and IT,\\
I. Kant Baltic Federal University, Nevskogo str. 14, 236041
Kaliningrad, Russia \\
$^{2}$Department of Mathematics, BITS Pilani K K Birla Goa Campus, India\\
$^{3}$Institut de Ci\'{e}ncies de l'Espai, ICE/CSIC-IEEC, Campus UAB,\\
Carrer de Can Magrans s/n, 08193 Bellaterra (Barcelona), Spain \\
$^{4}$Instituci\'{o} Catalana de Recerca i Estudis Avan\c{c}ats
(ICREA), Barcelona, Spain\\ \email{$^{+}$AAstasheok@kantiana.ru}}

\maketitle

\begin{history}
\received{21 November 2018}

\accepted{23 December 2018}
\end{history}

% 04.50.Kd  Modified theories of gravity
% 98.80.-k   Cosmology (see also section 04 General relativity and gravitation; for origin and evolution of galaxies, see 98.62.Ai; for elementary particle and nuclear processes, see 95.30.Cq; for dark matter, see 95.35.+d; for dark energy, see 95.36.+x; for superclusters and large-scale structure of the Universe, see 98.65.Dx)
% 98.80.Cq Particle-theory and field-theory models of the early Universe (including cosmic pancakes, cosmic strings, chaotic phenomena, inflationary universe, etc.)
% 12.60.-i Models beyond the standard model

\begin{abstract}
We compare the gravitational collapse of homogeneous perfect fluid
with various equations of state in the framework of General
Relativity and in $R^2$ gravity. We make our calculations using
dimensionless time with characteristic timescale $t_{g}\sim
(G\rho)^{-1/2}$ where $\rho$ is a density of collapsing matter.
The cases of matter, radiation and stiff matter are considered. We
also account the possible existence of vacuum energy and its
influence on gravitational collapse. In a case of $R^2$ gravity we
have additional degree of freedom for initial conditions of
collapse. For barotropic equation of state $p=w\rho$ the result
depends from the value of parameter $w$: for $w>1/3$ the collapse
occurs slowly in comparison with General Relativity while for
$w<1/3$ we have opposite situation. Vacuum energy as expected
slows down the rate of collapse and for some critical density
gravitational contraction may change to expansion. It is
interesting to note that for General Relativity such expansion is
impossible. We also consider the collapse in the presence of
so-called phantom energy. For description of phantom energy we use
Lagrangian in the form $-X-V$ (where $X$ and $V$ are the kinetic
and potential energy of the field respectively) and consider the
corresponding Klein-Gordon equation for phantom scalar field.
\end{abstract}


\section{Introduction}

Gravitational collapse has become a major topic of research
recently. It is believed to be the basic mechanism for the
structure formation of our universe. The problem of a star
contracting under the influence of its own gravitational field was
firstly studied by J.R. Oppenheimer and his student H. Snyder in
1939 \cite{oppenheimer}, some time after Einstein developed the
idea of General Relativity. They solved the Einstein's field
equation for the case of a star consisting of pressureless fluid
having uniform density and collapsing to a black hole. By using
the two matching conditions, they tried to answer whether it is
possible to mantain smoothness of the metric at the common
boundary. Using the condition of smoothness allows to describe the
behavior of the components of the metric. Then gravitational
collapse in frames of General Relativity has been studied in many
papers (see \cite{Weinberg} and references therein). For simple
case of dust matter with zero pressure it is possible to obtain
that time of collapse is $\sim \rho_{0}^{-1/2}$ where $\rho_{0}$
is initial density of collapsing object. For more realistic case
of collapsing star one needs to know the equation of state for
dense matter.

It is necessary to note that General Relativity is very successful
theory \cite{Weinberg, Schutz, Wald}. The solar system tests like
the precession of the perihelion of Mercury, bending of light in
gravitational field of sun and the gravitational red-shift are all
in agreement with the prediction of General Relativity.
Gravitational waves, which were first proposed by Henri Poincare
and predicted by General Relativity were recently observed by LIGO
collaboration \cite{LIGO}.

In spite of this achievements there have been speculations among
the scientific community regarding of the validity of the General
Relativity. Motivations to doubt this theory come from its
inability to explain certain phenomena, like the inflation
proposed by Alan Guth in 1979 \cite{Liddle, Guth}, which happened
soon after the Big Bang, and the late cosmic acceleration observed
by Perlmutter et al. \cite{Perlmutter} and Riess et al.
\cite{Riess} in 1998. This acceleration cannot be explained in
General Relativity with usual matter sources such as dust matter
and radiation. One needs to postulate existence of so called dark
energy i. e. a matter field having the equation of state parameter
$w< -\frac{1}{3}$. Such fluid can lead to repulsing effect and
corresponding acceleration. But physical nature of dark energy is
unclear. The first tentative candidate of dark energy having the
potential to explain the accelerated expansion of the universe is
the cosmological constant $\Lambda$ (nonzero vacuum
energy)\cite{Einstein, Kowalski, Planck16}. However, such
explanation contains the problem related with the fine tuning
problem which is not yet resolved \cite{Weinberg89, Linde16}.
Other candidates of dark energy like k-essence, holographic
energy, phantom energy, chameleon scalar fields have also been
explored in literature \cite{Caldwell, Steinhardt, Carroll2,
Singh, Sami}.

{An alternative approach to explain the cosmic acceleration is to
reformulate the theory of gravity such that it could provide an
explanation of observational data without any exotic dark energy
components. Of course this modification is restricted from
observational and local gravity constraints. The most simple
modification of gravity is to include the functions of the scalar
curvature in the gravitational action ($f(R)$ theories of modified
gravity) (for review see \cite{Odintsov2011, Capozziello2010,
Capozziello2011, Cruz2012, Odintsov2017}). These theories in
principle have the capability to explain the acceleration era
without the existence of exotic matter fields.  Other models of
modified gravitational theory include $f(R,T)$ gravity
\cite{Harko1}, scalar-tensor theories \cite{Uzan, chiba},
Gauss-Bonnet gravity \cite{nojiri}, \cite{nojiri-2} etc, each
having its own merits and demerits being widely discussed in
literature.}

Gravitational collapse in frames of General Relativity has been
recently investigated in many papers \cite{Goldwirth, Shapiro,
Dadhich, Giambo, Goswami, Ganguly, Joshi, Baier, Sami17}. In a
case of modified gravity interesting results are obtained in
\cite{Sharif, Cembranos, Santos, Ghosh, Bamba, Arbuzova, Amir,
Bonanno}. For example the authors of \cite{Sharif} obtained that
constant scalar curvature term in the action led to slowdown of
collapse. In \cite{Cembranos} the general $f(R)$ model is analyzed
for uniformly collapsing cloud of self-gravitating dust particles.
According to calculations for viable $f(R)$ models we have initial
epoch with higher contraction than in General Relativity. E.
Santos \cite{Santos} paid the attention on theoretical arguments
for the collapse of massive stars, which as expected is
unavoidable in General Relativity. As shown this is not necessary
for the theory with Einstein-Hilbert action involving a function
of $R^2-R_{\mu\nu}R^{\mu\nu}/2$ added to the Ricci scalar. Authors
of \cite{Ghosh} found exact nonstatic dust solutions in metric
$f(R)$ gravity, imposed by the constant scalar curvature and
Yang-Mills gauge theory, which describes the gravitational
collapse of presureless dust in (anti-)de Sitter
higher-dimensional background. The interesting question about
curvature singularities in $f(R)$ gravity is considered in
\cite{Bamba}. Authors investigated a curvature singularity
appearing in the star collapse process and concluded that addition
of term $\sim R^{\alpha}$ ($1<\alpha<2$) could cure the curvature
singularity.

Our primary purpose is to compare process of the collapse in
General Relativity and in $R^2$ gravity. Paper is organized as
follows. In section 2, we recall the description of collapse in
frames of General Relativity. Three different types of fluid (dust
matter, radiation and stiff matter) are considered. We study also
the gravitational collapse in the presence of vacuum energy. In
our calculations the dimensionless time is used for simplicity.
The characteristic gravitational time-scale is $\sim
(G\rho)^{-1/2}$ where $\rho$ is initial energy density of
collapsing fluid. In the next section the influence of phantom
dark energy on collapse is studied. We considered dynamical
equation for the scale factor and Klein-Gordon equation for
phantom scalar field. Then we discuss the collapse of the perfect
fluid in the presence of a scalar field dark energy model called
quintessence. Section 4 is devoted to study the collapse in frames
of $f(R)=R+\beta R^2$ gravity. Lastly, we end the paper with
conclusions derived from the calculations in the previous
sections.

\section{Gravitational collapse in General Relativity}

{We intend to study the gravitational collapse of homogeneous and
isotropic fluid having a spherically symmetric
Lemaitre-Tolman-Bondi metric (whose components are separable)
given by}
\begin{equation}
ds^{2}=-dt^2+A^{2}(t)h(r)dr^{2}+A^{2}(t)r^{2}d\Omega^{2}.
\end{equation}
The collapsing fluid is taken as a perfect fluid with
stress-energy tensor $T_{\mu}^{\nu}=\mbox{diag}(-\rho, p, p, p)$.
Hereafter we use system of units in which $G=c=1$. The components
$tt$, $rr$, $\theta\theta$ of the Einstein field equations are
written in terms of $A(t)$ and $h(r)$ as follows:
\begin{equation}\label{eq1}
3\frac{\ddot{A}}{A}=-8\pi \rho+\frac{R}{2},
\end{equation}
\begin{equation}\label{eq2}
\frac{\ddot{A}}{A}+2\frac{\dot{A}^{2}}{A^{2}}+\frac{h'}{A^{2}h^2r}=8\pi
 p+\frac{R}{2},
\end{equation}
\begin{equation}\label{eq3}
\frac{\ddot{A}}{A}+2\frac{\dot{A}^{2}}{A^{2}}+\frac{h'}{2A^{2}h^2r}-\frac{1}{A^{2}h
r^2}+\frac{1}{A^{2}r^{2}}=8\pi p+\frac{R}{2}.
\end{equation}
Here dot means time derivative {whereas} comma is derivative on
radial coordinate $r$. From last two equations one can derive the
following:
\begin{equation}
\frac{h'}{2 h^2r}+\frac{1}{h r^2}-\frac{1}{r^{2}}=0
\end{equation}
which is satisfied by
\begin{equation}
h(r)=\frac{1}{1+C_{1}r^{2}}.
\end{equation}
This gives us the Friedmann-Lemaitre-Robertson-Walker (FLRW)
metric
\begin{equation}
ds^{2}=-dt^{2}+A^{2}(t)\left(\frac{dr^{2}}{1+C_{1}r^{2}}+r^{2}d\Omega^{2}\right).
\end{equation}
In order to determine the collapse dynamics of the star, we will
have to find how the scale factor $A(t)$ behaves. Substituting for
$h(r)$ in Eq. (\ref{eq2}) or (\ref{eq3}), using Eq. (\ref{eq1}) to
eliminate the second derivative of scale factor and relation for
scalar curvature $R=-8\pi T=8\pi (\rho-3p)$ we obtain the
following first-order differential equation for $A(t)$:
\begin{equation}
\dot{A}^2= \frac{8\pi}{3}A^2 \rho+C_{1}.
\end{equation}
Without loss of generality one can assume that $A(0)=1$. For first
derivative one put $\dot{A}(0)=0$. For barotropic equation of
state in the form
$$
p=w\rho
$$
we have simple dependence of energy density from the scale factor
$$
\rho=\rho_{0}A^{-3(1+w)}.
$$
Here $\rho_{0}$ means the value of energy density at $t=0$. The
non-trivial solution for above mentioned conditions realizes for
$C_{1}=8\pi \rho_{0}$. Therefore for solution describing
gravitational collapse we have equation:
\begin{equation}\label{coll}
\dot{A}=-\sqrt{k(-1+A^{-1-3w})}, \quad k=\frac{8\pi}{3} \rho_{0}.
\end{equation}
Introducing dimensionless time according to relation
$$
\tau=\sqrt{k}t
$$
allows to rewrite the Eq. (\ref{coll}) in the following form:
\begin{equation}
\frac{dA}{d\tau}=-\sqrt{-1+A^{-1-3w}}.\label{10}
\end{equation}
For example for $w=0$ (dust) we have following solution for $\tau$
as function of scale factor:
$$
\tau=\frac{\pi}{2}-\arcsin\sqrt{A}+\sqrt{A-A^2}.
$$
The moment of singularity corresponds to $A=0$ and therefore we
have $\tau_{s}=\frac{\pi}{2}$. The radiation collapse corresponds
to $w=1/3$. In this case one can derive scale factor in the
explicit form:
$$
A(\tau)=\sqrt{1-\tau^2}.
$$
The moment of singularity corresponds to $\tau_{s}=1$. For stiff
fluid with $w=1$ we found that moment of singularity is
$$
\tau_{s}=\frac{1}{4}B\left(\frac{1}{2},\frac{3}{4}\right)\approx
0.6
$$
where $B$ means beta-function. The dynamics of collapse for
considered cases is presented on Fig. 1.

\begin{figure}
\begin{center}
\includegraphics[scale=1]{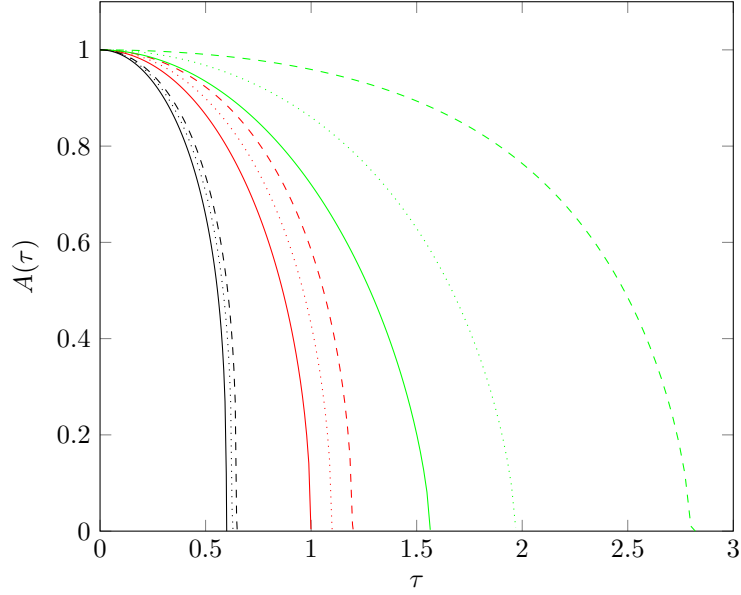}
\end{center}
\caption{Collapse dynamics of dust (green), radiation (red) and
stiff fluid (black) without cosmological constant (solid lines),
for $\Omega_\Lambda=0.2$ (dotted lines) and for
$\Omega_\Lambda=0.3$ (dashed lines) in the framework of General
Relativity.}
\end{figure}

Eq.(\ref{10}) corresponding to cosmological constant $\Lambda$
($w=-1$) is
\begin{equation} \label{c}
     \frac{dA}{d\tau}=-\sqrt{-1+A^{2}}.
\end{equation} \\
solving which we get
\begin{equation}
    A(\tau)=\cosh(\tau+C_{1}).
\end{equation}
Proceeding the same way as we did above, i.e. using $A(0)=1$, we
get  $C_1=0$ giving us:
\begin{equation}
     A(\tau)=\cosh(\tau).
\end{equation}
\par
We can conclude from the above equation that $A(\tau)$ can never
be $0$ at any finite time, which means that the cosmological
constant never collapses to form a singularity, as expected.

Finally we consider the case of fluid with barotropic EoS and
cosmological constant. The corresponding equation in dimensionless
units in this case is
\begin{equation}
\frac{dA}{d\tau}=-\sqrt{-\Omega_{m}^{-1}+\frac{\Omega_{\Lambda}}{\Omega_{m}}A^{2}+A^{-1-3w}}.
\end{equation}
Here $\Omega_{m}$ and $\Omega_{\Lambda}$ are fractions of energy
density of matter and cosmological constant correspondingly in the
moment $\tau=0$, i.e.
$$
\Omega_m=\frac{\rho_{0}}{\rho_{0}+\Lambda}, \quad
\Omega_{\Lambda}=\frac{\Lambda}{\rho_{0}+\Lambda}.
$$
The gravitational collapse due to repulsion of vacuum energy
occurs more slowly in comparison with the case of one fluid (see
Fig. 1).

\par
As suggested in the introduction, there have been attempts to
explain the current observations and the inflationary era by
either modifying the general theory of relativity or by adding
dark energy in general relativity. One such possible form of dark
energy is called phantom energy. We focus on investigating the
collapse of a gravitationally bound fluid in the framework of
General Relativity in presence of phantom energy.

\section{Phantom fluid}

The action for scalar field with non-canonical kinetic term is
given by
\begin{equation}
S=\int\left(\frac{R}{16\pi}+P(\phi,X)\right)\sqrt{-g}d^{4}x+S_{m}.
\end{equation}
Here $P(\phi,X)$ is an arbitrary function of scalar field $\phi$
and $X$, i.e. the kinetic energy of the field given by
$-\frac{1}{2}g^{\mu\nu}\partial_{\mu}\phi\partial_{\nu}\phi$. From
the above action, the energy-momentum tensor of the scalar field
can be evaluated as
\begin{equation}
T_{\mu\nu}^{(\phi)}=-\frac{2}{\sqrt{-g}}\frac{\delta\left(\sqrt{-g}P\right)}{\delta
g^{\mu\nu}}=P_{,X}\partial_{\mu}\phi\partial_{\nu}\phi+g_{\mu\nu}P.
\end{equation}
We consider the scalar field with energy-momentum tensor
$T_{\mu\nu}^{(\phi)}=(\rho_{\phi}+p_{\phi})u_{\mu}u_{\nu}+g_{\mu\nu}P$
giving us $p_{\phi}=P$ and $\rho_{\phi}=2XP_{,X}-P$. Hence we
obtain the equation of state parameter $w_{\phi}$ as
\begin{equation}\label{eosp1}
w_{\phi}=\frac{P}{2XP_{,X}-P}.
\end{equation}
Phantom energy is a particular case of k-essence in which
$w_{\phi}<-1$ which is obtained by imposing the condition
$P_{,X}<0$. The simplest model being a scalar field
$P(X,\phi)=-X-V(\phi)$ where $V(\phi)$ is the potential of scalar
field. We obtain in this case
$p_{\phi}=-\frac{\dot{\phi}^{2}}{2}-V(\phi)$ and
$\rho_{\phi}=-\frac{\dot{\phi}^{2}}{2}+V(\phi)$ {from} which we
obtain the equation of state parameter
\begin{equation}
w_{\phi}=\frac{\dot{\phi}^{2}-2V(\phi)}{\dot{\phi}^{2}+2V(\phi)}.
\end{equation}

When the potential energy dominates the kinetic energy, we get the
situation $w_{\phi}<-1$. Behavior of various potential has already
been studied and are available in literature \cite{Ratra, Zlatev,
Linde1, Linde2, Kallosh}, one of which is a linear model with
power law potential $V(\phi)=V_{0}\phi^{n}$. In the mathematical
tool which we are going to apply, the collapse dynamics becomes
independent of the coefficient of $\phi^{n}$ in the above
mentioned linear model with power law potential. In order to make
the coefficient of $\phi^{n}$ to play a significant role in
determining the astrophysical scenario, we propose a model of
phantom energy of the form $P(X,V)=-X-V$ having the slope of the
associated potential energy
$$
\frac{dV}{d\phi}=V_{0}(t)\phi^{n},\quad V_{0}(t)=V_{0}+\alpha
t^{m}.
$$

We consider the collapse scenario of an ordinary matter perfect
fluid in the presence of phantom scalar field. The Einstein's
field equations are:

\begin{equation}
\frac{\dot{A}^{2}}{A^{2}}=\frac{8\pi}{3}\left(\rho_{m}-\frac{\dot{\phi}^{2}}{2}+V(\phi)\right),
\end{equation}

\begin{equation}
\frac{\ddot{A}}{A}-\frac{\dot{A}^{2}}{A^{2}}=-4\pi\left(-\dot{\phi}^{2}+\rho_{m}+p_{m}\right).
\end{equation}

The equation of continuity for matter field is given by
\begin{equation}\label{pe3}
\dot{\rho}+3\frac{\dot{A}}{A}\left(\rho_{m}+p_{m}\right).
\end{equation}

For phantom field we obtain the Klein-Gordon equation
\begin{equation}\label{kg}
\ddot{\phi}+\frac{3\dot{A}}{A}\dot{\phi}-\frac{dV}{d\phi}=0.
\end{equation}

We use the integrability condition on anharmonic oscillator \cite{Harko}
which states that for the differential equation of the form:
\begin{equation}\label{anharmonic}
\ddot{x}+f_{1}(t)\dot{x}+f_{2}(t)x+f_{3}(t)x^{n}=f_{4}(t)
\end{equation}
which governs the time evolution of the space variable $x(t)$ of
an anharmonic oscillator (where $x$ and $f_{k}(t)$ are
continuously differentiable real functions defined on some
interval), we have the following statement for $n\neq -3,-1,0,1$:

The coefficients of Eq.(\ref{anharmonic}) satisfy the differential
equation:
\begin{equation}\label{de}
\frac{\ddot{f}_{3}}{(n+3)f_{3}}-\frac{n+4}{(n+3)^{2}}\frac{\dot{f}_{3}^{2}}{f_{3}^{2}}+\frac{n-1}{(n+3)^{2}}\frac{\dot{f}_{3}f_{1}}{f_{3}}+\frac{2\dot{f}_{1}}{n+3}+
\frac{2(n+1)f_{1}^{2}}{(n+3)^{2}}=f_{2}
\end{equation}
if and only if Eq.(\ref{anharmonic}) can be transformed into an
integrable form
\begin{equation}\label{tde}
\ddot{X}(T)+X^{n}(T)=0.
\end{equation}
where
\begin{equation}
X(T)=Cx(t)f_{3}^{\frac{1}{n+3}}(t)e^{\frac{2}{n+3}\int^{t}f_{1}(\eta)d\eta},\label{t1}
\end{equation}
\begin{equation}\label{t2}
T(t)=C^{\frac{1-n}{2}}\int^{t}f_{3}^{\frac{2}{n+3}}(\xi)e^{\frac{1-n}{n+3}\int^{\xi}f_{1}(\eta)d\eta}d\xi.
\end{equation}

Comparing Eq.(\ref{anharmonic}) with the Klein-Gordon
Eq.(\ref{kg}) corresponding to the phantom field having slope of
the potential energy $\frac{dV}{d\phi}=(V_0+\alpha t^m)\phi^n$, we
have $f_1=\frac{3\dot A}{A}$, $f_2=0$ and $f_3=-(V_0+\alpha t^m)$,
for which the Eq.(\ref{de}) becomes
    \begin{equation} \label{eqs}
    \begin{split}
       & \frac{1}{n+3}\frac{\alpha m (m-1)t^{m-2}}{V_0+\alpha t^m}-\frac{n+4}{(n+3)^2}\left(\frac{m \alpha t^{m-1}}{V_0+\alpha t^m} \right)^2+\\
             & +\frac{n-1}{(n+3)^2}\left(\frac{\alpha m t^{m-1}}{V_0+\alpha t^m}\right)\frac{3\dot A}{A}+\frac{6}{n+3} \left(\frac{\ddot A}{A}-\frac{\dot A^2}{A^2}\right) +\frac{18(n+1)}{(n+3)^2}\frac{\dot A^2}{A^2}=0.
    \end{split}
    \end{equation}
    For simplicity, we consider a particular case corresponding to $m=1$. The Eq.(\ref{de}) becomes
    \begin{equation}
         \begin{split}
-\frac{n+4}{n+3}\left(\frac{ \alpha }{V_0+\alpha t}
\right)^2+\frac{n-1}{n+3}\left(\frac{\alpha}{V_0+\alpha t} \right)
\frac{3\dot A}{A}+ \frac{6\ddot A}{A}
+\frac{12n}{n+3}\frac{\dot A^2}{A^2}=0.
    \end{split}
    \end{equation}

On solving the above differential equation, we get the equation of
scale factor governing the collapse of the fluid
\begin{equation}\label{phan}
A(t)=C_{2}\left(V_0+\alpha t\right)^{-1/6}\left|1+C_1(V_0+\alpha
t)^{3/2}\right|^{\frac{n+3}{3(n+1)}}.
\end{equation}

In the above equation, we have unknown constants $C_1$,
$C_2$ for given parameters $V_0$, $\alpha $ and $n$. We use the initial conditions
$A(0)=1$ and $\dot{A}(0)=0$ and get the following values for $C_1$
and $C_2$:
$$
C_{1}=\frac{{V_0^{-\frac{3}{2}}}(n+1)}{2(n+4)}, \quad
C_{2}=V_{0}^{1/6}\left|1+C_1
V_0^{3/2}\right|^{-\frac{n+3}{3(n+1)}}.
$$

For $-3<n<-1$ value of $C_1$ lies between -1 and 0. Therefore $A$
diverges at some moment of time (backward collapse). From Eq.
(\ref{phan}) one can see that for $n>-1$ and $n<-3$ on late times
\begin{equation}
A\sim t^{\frac{n+4}{3(n+1)}},\quad t\rightarrow\infty
\end{equation}

\begin{figure}\begin{center}
\includegraphics[scale=0.6]{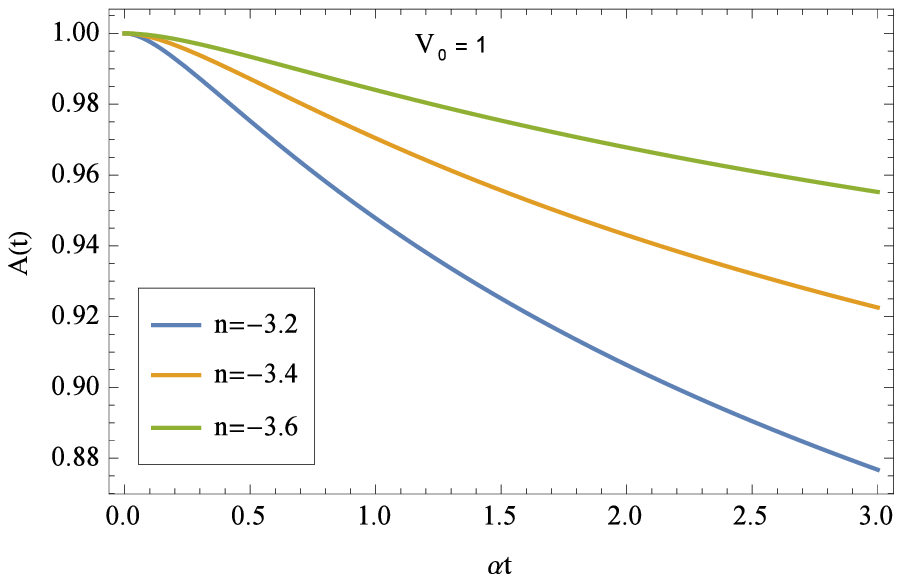}\includegraphics[scale=0.6]{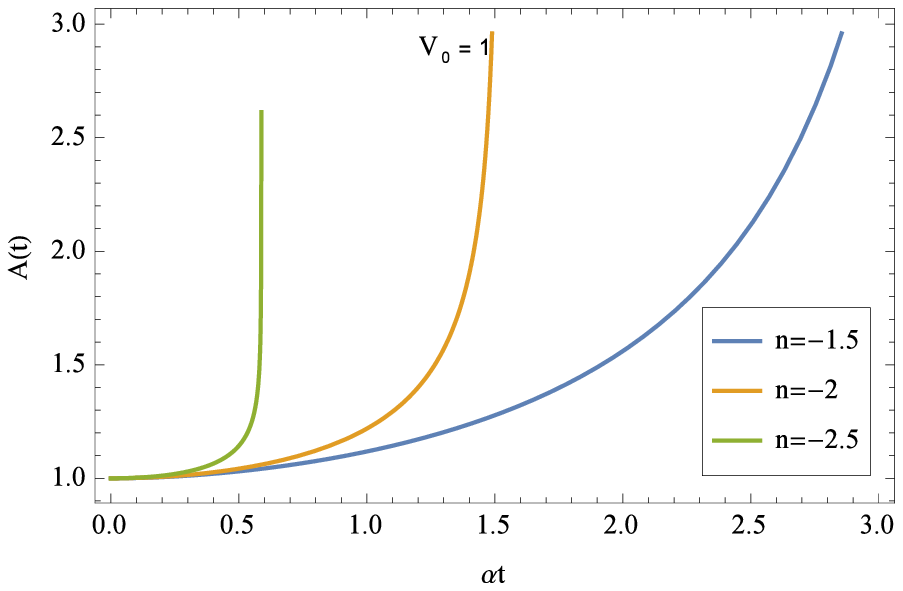}\\
\includegraphics[scale=0.6]{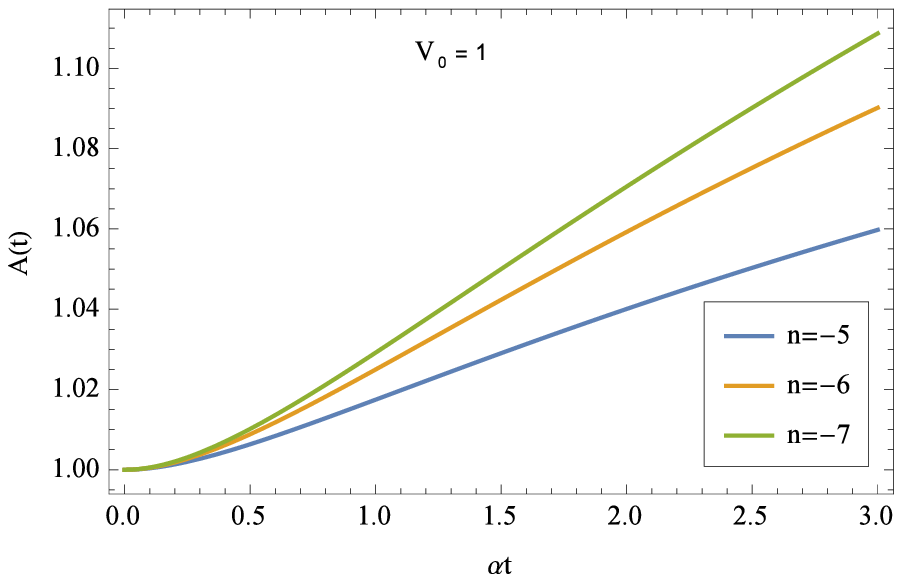}\includegraphics[scale=0.6]{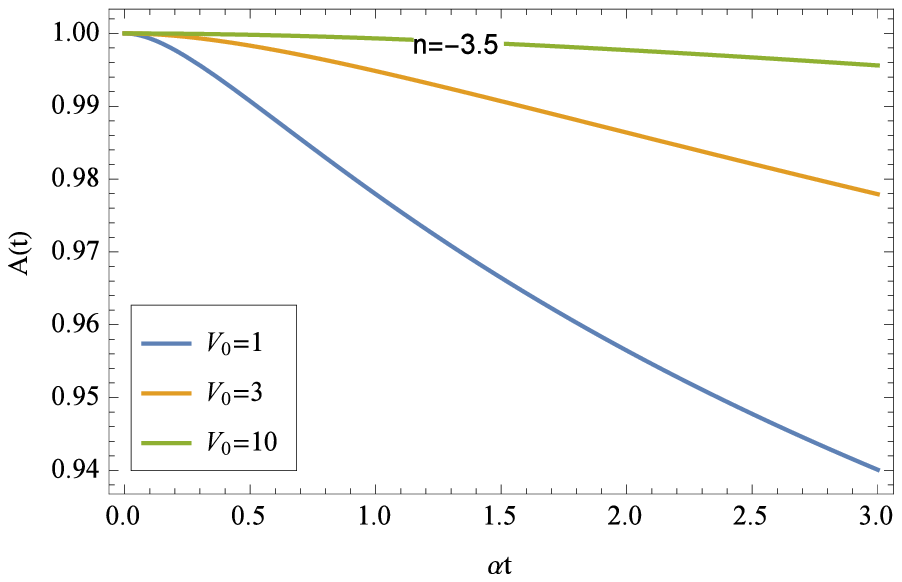}\\
\includegraphics[scale=0.6]{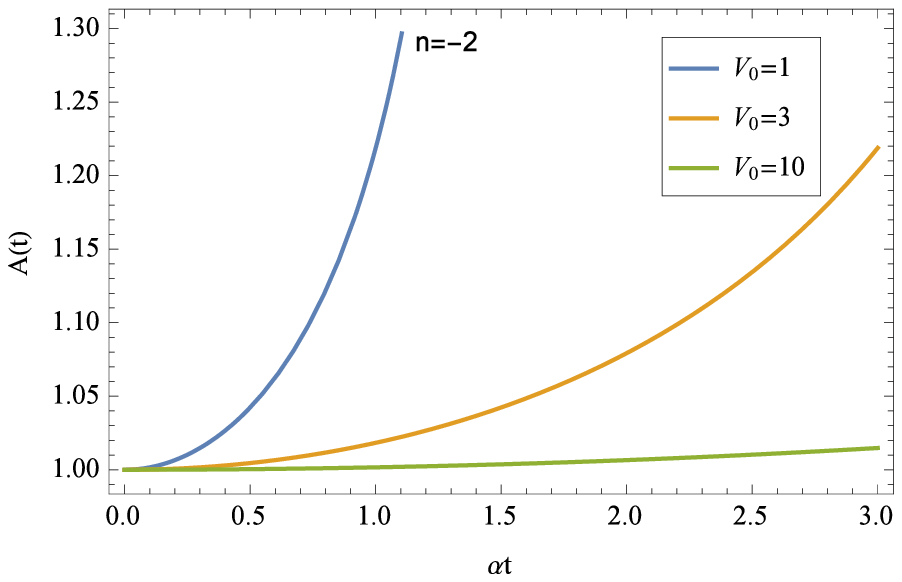}\includegraphics[scale=0.6]{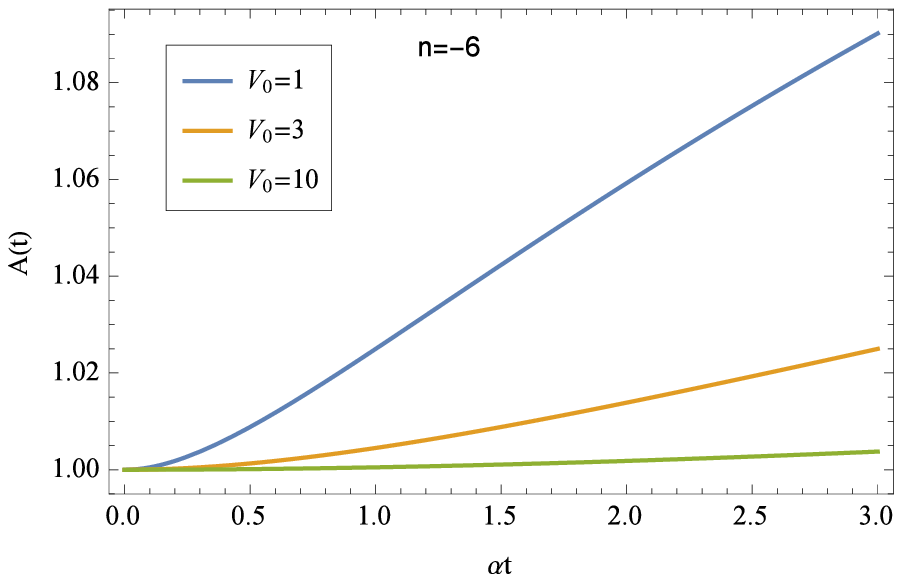}
\end{center}
\caption{Dynamics of scale factor $A(\tau)$ in presence of phantom
energy, having Lagrangian $P(X,\phi)=-X-V(\phi)$ where the
potential has the slope
$\frac{dV(\phi)}{d\phi}=({V_0}+{\alpha}t)\phi ^n)$, for various
values of $n$ {and $V_0$}.} \label{sf1}
\end{figure}

From this asymptotics one can see that in the presence of phantom
energy field having slope of the potential  of the form
$\frac{dV}{d\phi}=(V_0+ \alpha t)\phi^n$, the fluid collapses for
$-4<n<-3$. The collapse becomes faster with increasing of value of
$n$. For a given $n$ in the above mentioned domain, increasing
$V_0$ slows down the collapse, while increasing $\alpha$ fastens
the process.

For $n<-4$ and $n>-1$ there is no collapse.  This could be
interpreted as if the dark energy has become so dominant that it
doesn't allow any inward gravitational pull at all. For these
cases increasing of $n$ reduces the expansion rate. Increasing
$V_0$ ($\alpha$) slows down (speeds up) the expansion of the
fluid. For $n$ close to $-1$ from above, the phantom field becomes
too strong. For example for $V_0=\alpha=1$ and for $n=-0.99$, the
scale factor reaches up to $3.5\times 10^{107}$ for time $t=10^3$.
Comparing it with the outcome for $n=5$, the scale factor
increases $18$ times for the same time period. Dependence of the
expansion of the fluid from the values $V_0$ and $\alpha$ is
similar to the previous case.

If we consider a more general dark energy model of similar kind
having slope of the form $\frac{dV}{d\phi}=(V_0 +\alpha t^m)\phi
^n$, the equation of state parameter $w$ is given by
Eq.(\ref{eosp1}). Analysis shows that for positive $m$ the
gravitational collapse is expected to slow down and if there is
expansion, then it speeds up.

\subsection{Comment on quintessence}

As is discussed in the introduction, an alternative to the
cosmological constant is necessary in order to avoid the
coincidence problem. The real candidate for dark energy is scalar
field called quintessence. Unlike the cosmological constant case,
there is no restriction as to how small the energy density need to
be with respect to matter or radiation density for the very early
phase of the universe.

The action for a canonical scalar field, called the quintessence,
having potential $V(\phi)$ is given by
\begin{equation} \label{actionq}
    S= \int \left (\frac{R}{16\pi}+ X-V(\phi) \right) \sqrt{-g} d^4x + S_m
\end{equation}
where $X=-\frac{1}{2}g^{\mu \nu}\partial_\mu \partial_\mu \phi$,
i.e. the kinetic energy of the scalar field. From the above
action, the energy-momentum tensor of the scalar field can be
evaluated as
\begin{equation}
 T^{(\phi)}_{\mu \nu}= \partial _{\mu} \phi \partial _{\nu} \phi -g_{\mu \nu} \left(\frac{1}{2}g^{\alpha \beta}\partial_{\alpha}\phi\partial_\beta\phi+V(\phi)
 \right),
\end{equation}
We consider the scalar field having the behavior of a perfect
fluid following the equation $T^{\phi}_{\mu \nu}=(\rho _{\phi}+
P_{\phi})u_{\mu} u_{\nu} + g_{\mu \nu} P$, giving us
$P_{\phi}=\frac{1}{2}\dot \phi^2+V(\phi)$ and $\rho_{\phi}=
\frac{1}{2}\dot \phi^2-V(\phi)$. Hence we obtain the equation of
state parameter $w$ as
\begin{equation}
    w_{\phi}= \frac{\dot \phi^2-2V(\phi)}{\dot \phi^2+2V(\phi)}.
\end{equation}
We consider the slope in the potential as
$\frac{dV}{d\phi}=V_0(t)\phi^n $ where $V_{0}(t)= V_0+\alpha t^m$.
    Using the field equations corresponding to action Eq.(\ref{actionq})
    and the equation of continuity for matter field Eq.(\ref{pe3}), we obtain the Klein-Gordon equation for quintessence given by
    \begin{equation} \label{kg1}
        \ddot \phi +\frac{3\dot A}{A}\dot \phi+\frac{dV}{d\phi}=0.
    \end{equation}
Using the integrability condition, we obtain a differential
equation of $A(t)$ given by Eq.(\ref{eqs}). We can conclude that
the equations which govern the dynamics of the fluid in the
presence of quintessence is same as that in presence of phantom
energy for the case where the slope of the potential is given by
$V_0+\alpha t^m$.

\section{Gravitational collapse in $f(R)=R+\beta R^2$ gravity}

Let's consider the gravitational theory with action for
gravitational field
\begin{equation}\label{fr}
S_{g}=\int\frac{f(R)}{16\pi}\sqrt{-g}d^{4}x ,
\end{equation}
where $f(R)$ is arbitrary differentiable function of Ricci scalar
$R$. For our purposes we write function $f(R)$ as
$$
f(R)=R+h(R),
$$
extracting contributions with respect to GR, $h(R)$. The field
equations obtained from Eq.(\ref{fr}) are
\begin{equation}\label{field}
(1+h_{R})G_{\mu \nu }-\frac{1}{2}(h-h_{R}R)g_{\mu \nu }-(\nabla
_{\mu }\nabla _{\nu }-g_{\mu \nu }\Box )h_{R}=8\pi T_{\mu \nu }.
\end{equation}
Here $G_{\mu\nu}=R_{\mu\nu}-\frac12Rg_{\mu\nu}$ is the Einstein
tensor and ${\displaystyle h_R=\frac{dh}{dR}}$,
$\Box=g^{\mu\nu}\triangledown_{\mu}\triangledown_{\nu}$ is the
covariant D'Alamber {operator}. For the Ricci curvature scalar one
can get the following equation:
\begin{equation}
3h_{RR}\Box R=8\pi T-3h_{RRR}(\triangledown R)^{2}+R+2h-Rh_{R}
\end{equation}
For FLRW metric with positive curvature we have following
equations for $A(t)$:
\begin{equation}\label{fr1}
H^{2}-\frac{C_{1}}{A^2}=\frac{8\pi\rho}{3(1+h_{R})}+\frac{1}{3(1+h_{R})}\left(\frac{1}{2}h_{R}R-3h_{RR}H\dot{R}\right).
\end{equation}
Finally for Ricci scalar equation can be rewritten as:
\begin{equation}\label{fr2}
3h_{RR}\left(\ddot{R}+3H\dot{R}\right)=-3h_{RRR}\dot{R}^{2}+8\pi
T+2h-h_{R}R.
\end{equation}
Here $H=\frac{\dot{A}}{A}$. From metric one can obtain the
following relation for Ricci scalar
\begin{equation}\label{fr3}
R=6\left(\dot{H}+2H^{2}-\frac{C_{1}}{A^{2}}\right).
\end{equation}
The equation of continuity remains the same:
\begin{equation}\label{fr4}
\dot{\rho}+3H\left(\rho+p\right)=0.
\end{equation}
Scalar curvature depends from second derivatives of metric
$g_{\mu\nu}$ and therefore from fourth derivative of $A(t)$.
However one can solve equation for scalar curvature as independent
variable with equations for $A$ and $\rho$. We use equations
(\ref{fr1})-(\ref{fr4}) with definition for $H$ and integrate it
for some initial conditions $\rho_{0}$, $R_{0}$, $\dot{R}_{0}$,
$A_{0}$, $H_{0}$. Note that equation (\ref{fr1}) is consistent
with another equations and therefore give no additional
information. But this equation can be used to fix initial data.
For example for given $H(0)$, $\rho(0)$, $A(0)$ and $\dot{R}(0)$
one can derive the condition on scalar curvature at $t=0$. In
dimensionless units we put $C_{1}=-8\pi\rho_{0}/3$ and rewrite
(\ref{fr1}), (\ref{fr2}), (\ref{fr3}) in following form:
\begin{equation}\label{fr1-1}
H^{2}=-\frac{1}{A^{2}}+\frac{\rho}{(1+h_{R})}+\frac{1}{(1+h_{R})}\left(\frac{1}{2}h_{R}R-h_{RR}H\dot{R}\right),
\end{equation}
\begin{equation}\label{fr2-1}
3h_{RR}\left(\ddot{R}+3H\dot{R}\right)=-3h_{RRR}\dot{R}^{2}+3
T+3R+3(2h-h_{R}R),
\end{equation}
\begin{equation}\label{fr3-1}
R=2\left(\dot{H}+2H^{2}+\frac{1}{A^{2}}\right).
\end{equation}
Here we adopted the following dimensionless units for density,
pressure, time and curvature:
$$
\rho\rightarrow\rho \rho_{0},\quad p\rightarrow p \rho_{0},
$$
$$
t\rightarrow\tau\left(\frac{3}{8\pi\rho_{0}}\right)^{1/2}, \quad
R\rightarrow 8\pi\rho_{0}R, \quad h\rightarrow h 8\pi\rho_{0}.
$$

Let's consider simple model of $f(R)$ with $h=\beta R^2$ where
$\beta$ is constant parameter ($R^{2}$ gravity)
\cite{Odintsov2011}. One can consider the various cases of initial
conditions. Let's choose that $H(0)=0$, $\rho(0)=1$ and
$\dot{R}(0)=0$. For given $A(0)$ from (\ref{fr1-1}) one can define
initial condition for scalar curvature $R_{0}$. Then integration
of (\ref{fr2-1}), (\ref{fr3-1}) with equation of continuity gives
the solution for unknown functions. We consider especially case
$A(0)=1$. For scalar curvature we have that $R(0)=0$. Results for
$A$ are given on Fig. 3 for matter ($w=0$) and stiff matter
($w=1$). For radiation ($w=1/3$) at given initial conditions we
have no significant difference from General Relativity (scalar
curvature is zero as in General Relativity for radiation). The
time of collapse increases with $\beta$ for stiff matter and
decreases for dust matter. For $\beta>\sim 0.1$ in dimensionless
units (this value corresponds to $(8\pi\rho_{0})^{-1}$)
$\tau\rightarrow 1$ for both cases. Varying equation of state
parameter $w$ allows to conclude that for $0\leq w<1/3$ collapsing
time decreases with $\beta$.

\begin{figure}\begin{center}
\includegraphics[scale=0.6]{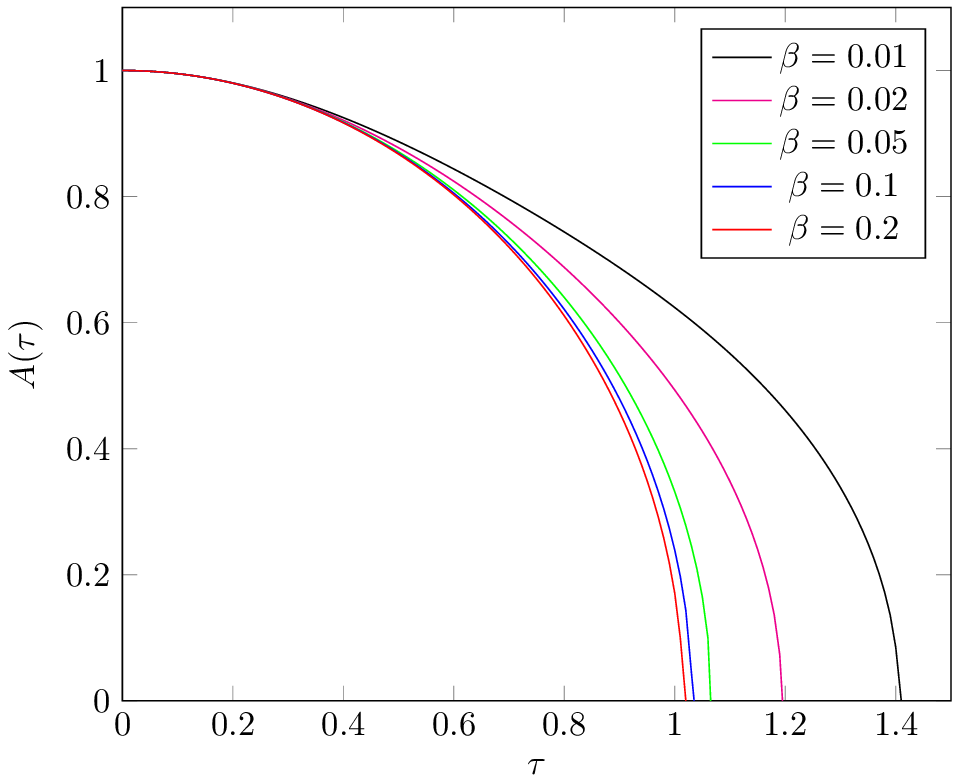}\includegraphics[scale=0.6]{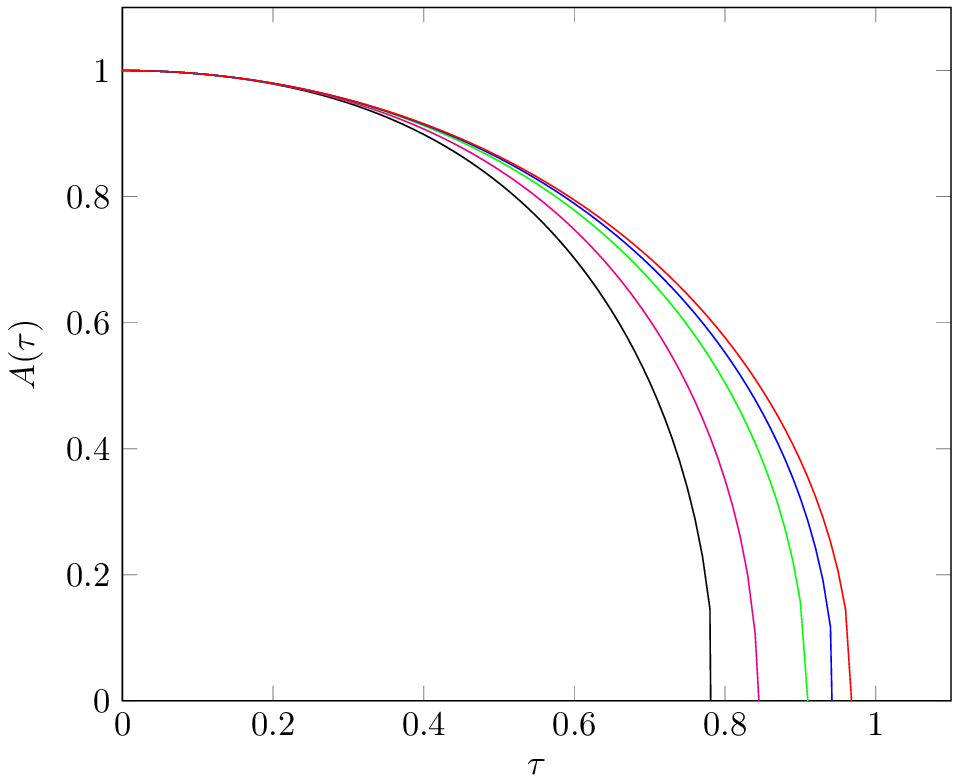}\\
\end{center}
\caption{Dynamics of scale factor $A(\tau)$ for dust (left panel)
and stiff matter (right panel) in frames of $R^2$ gravity for
various $\beta$ and initial conditions $A(0)=1$, $H(0)=0$,
$R(0)=0$, $\dot{R}(0)=0$, $\rho(0)=1$.} \label{sf2}
\end{figure}

The next step is to add into consideration the lambda-term and
study its influence on process of collapse. Evolution in this case
depends from the fraction of vacuum energy (see Fig. 4). For some
$\Omega^{crit}_{\Lambda}$ initial contraction turns into
expansion. The value of $\Omega^{crit}_{\Lambda}$ depends from
$\beta$ and equation of state. One notes that in General
Relativity the vacuum energy only slows down gravitational
collapse and contraction never turns into expansion.

\begin{figure}\begin{center}
\includegraphics[scale=0.6]{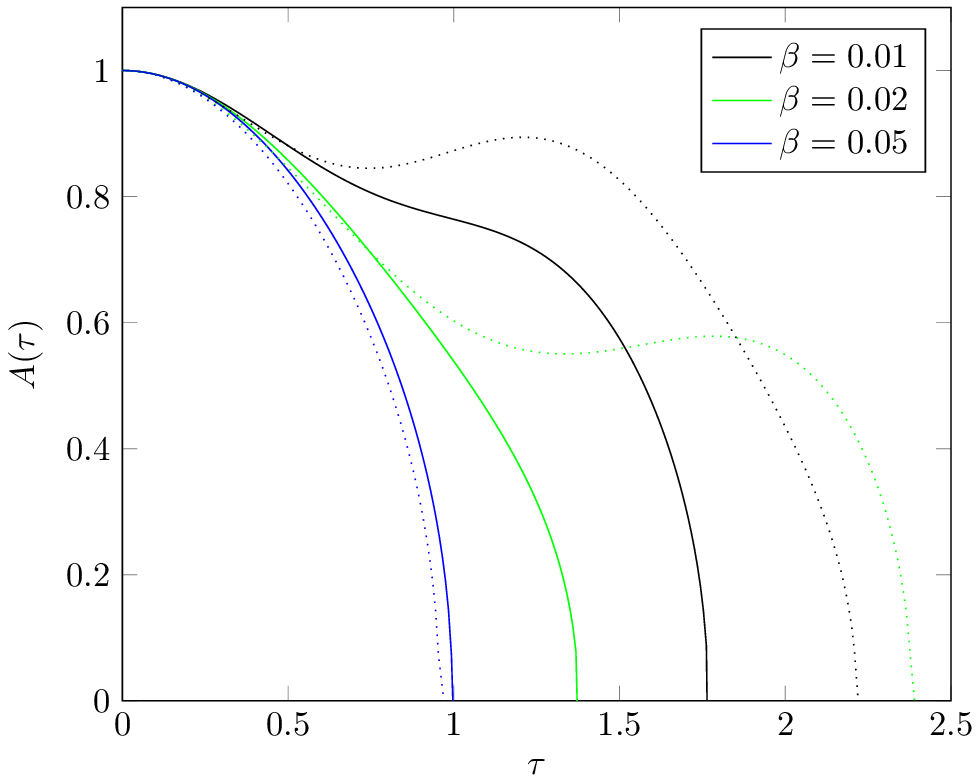}\includegraphics[scale=0.6]{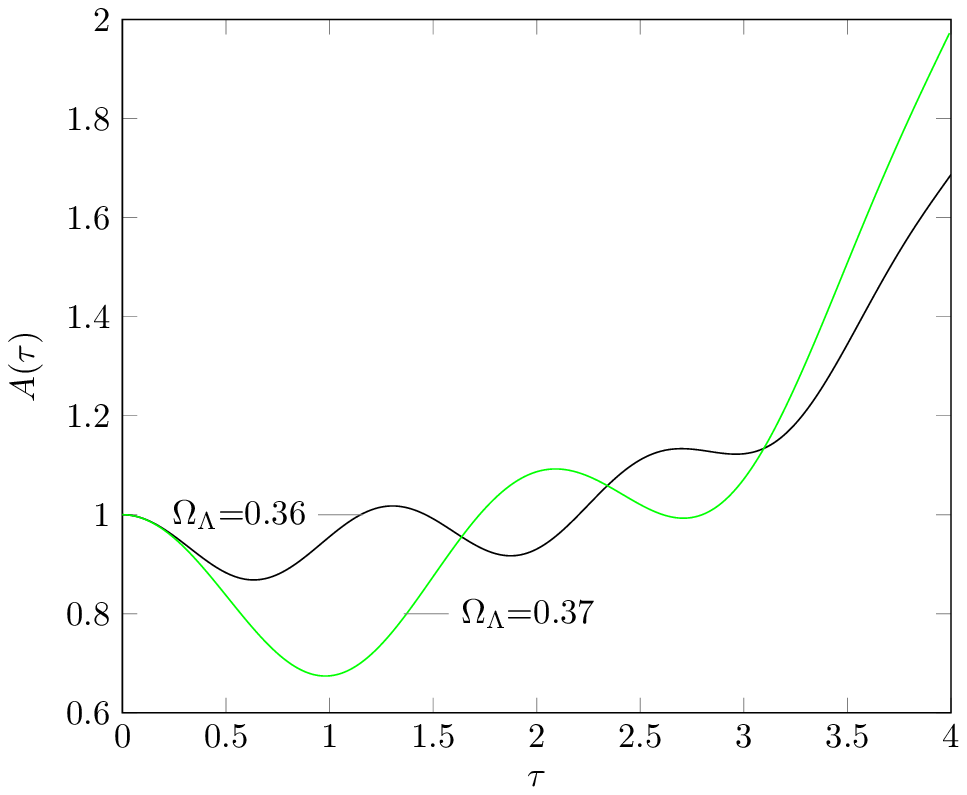}\\
\includegraphics[scale=0.6]{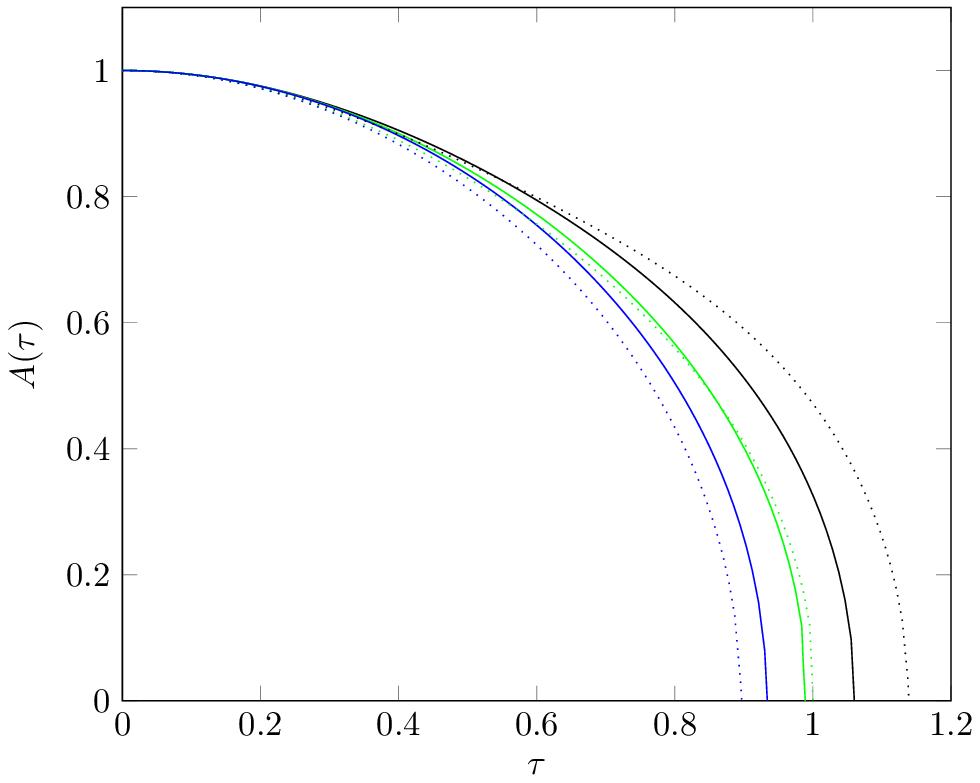}\includegraphics[scale=0.6]{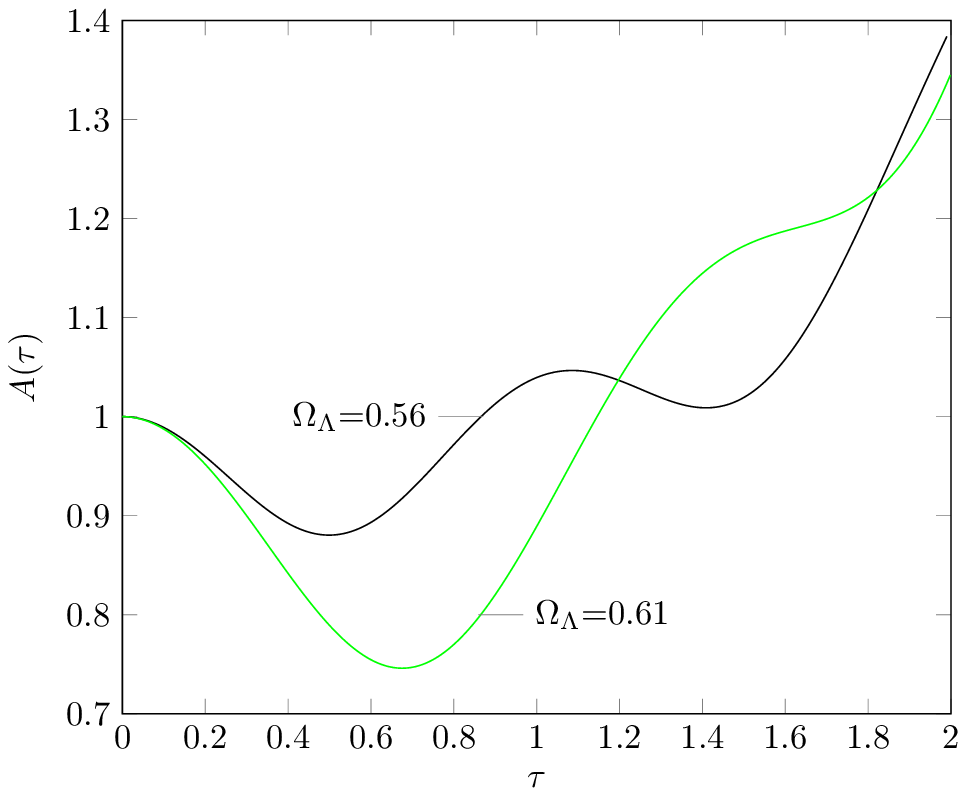}\\
\includegraphics[scale=0.6]{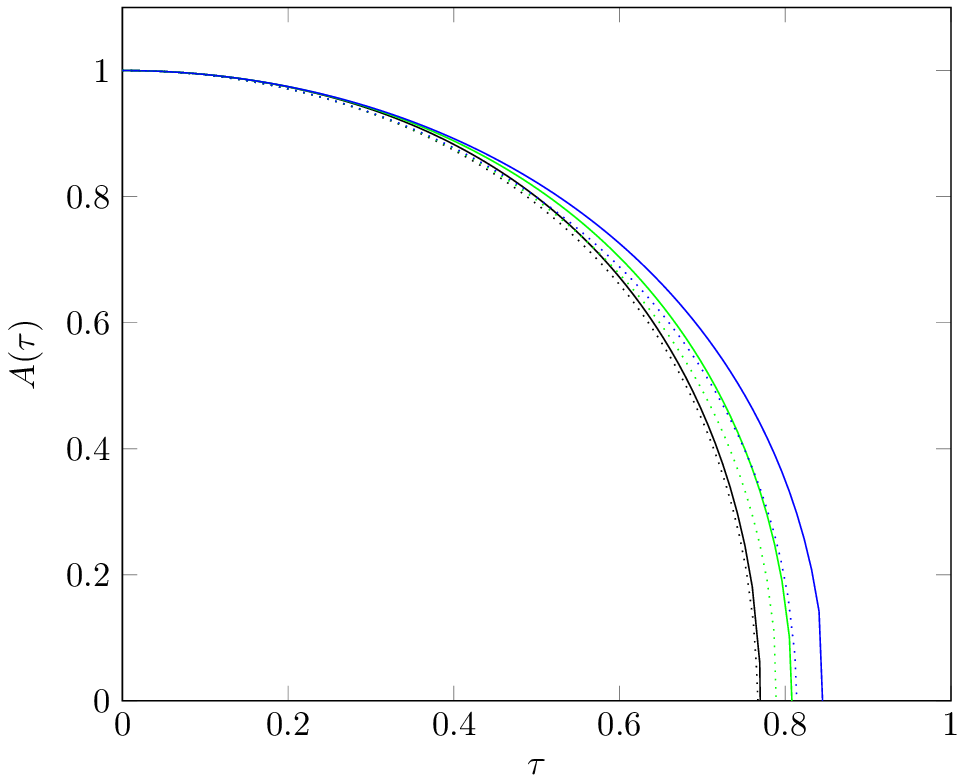}\includegraphics[scale=0.6]{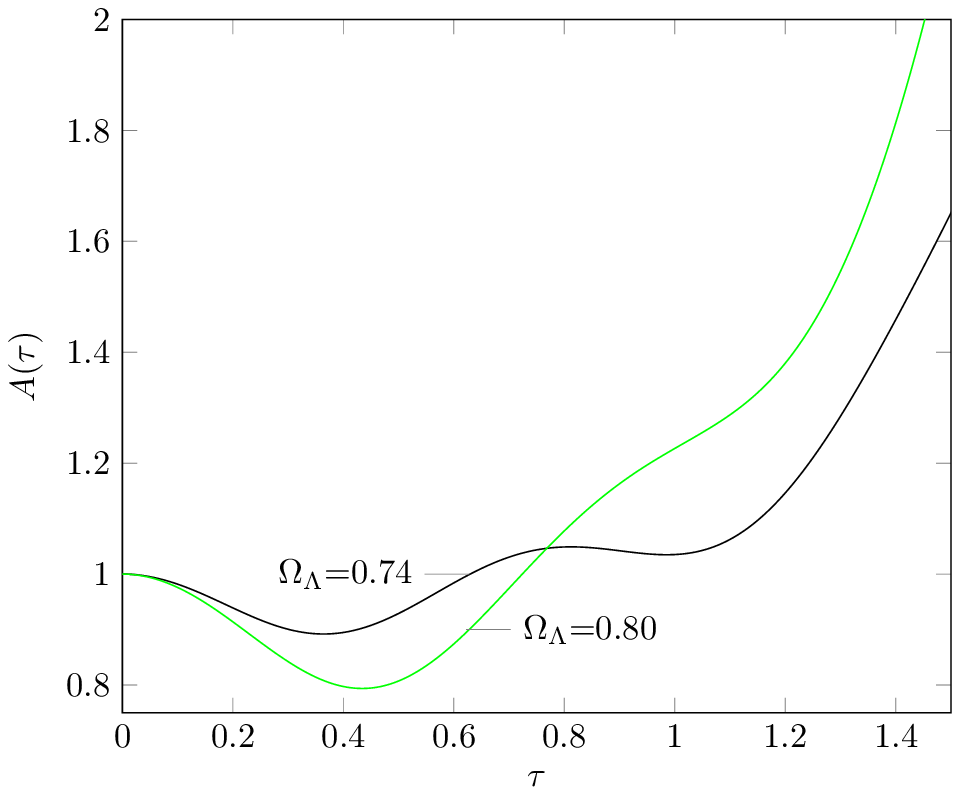}\\
\end{center}
\caption{Dynamics of scale factor $A(\tau)$ for dust matter (upper
panel), radiation (middle panel) and stiff matter (down panel) in
frames of $R^2$ gravity  for various $\beta$ and fraction of
$\Lambda$. Initial conditions are $H(0)=0$, $\dot{R}(0)=0$,
$\rho_m(0)=1$. On left side solid and dotted lines correspond to
$\Omega_{\Lambda}=0.2$ and $\Omega_{\Lambda}=0.3$. On right side
the process of evolution is depicted for $\Omega^{crit}_{\Lambda}$
at which expansion take place (for $\beta=0.01$ and $0.02$ only).}
\label{sf3}
\end{figure}

\begin{figure}\begin{center}
\includegraphics[scale=0.6]{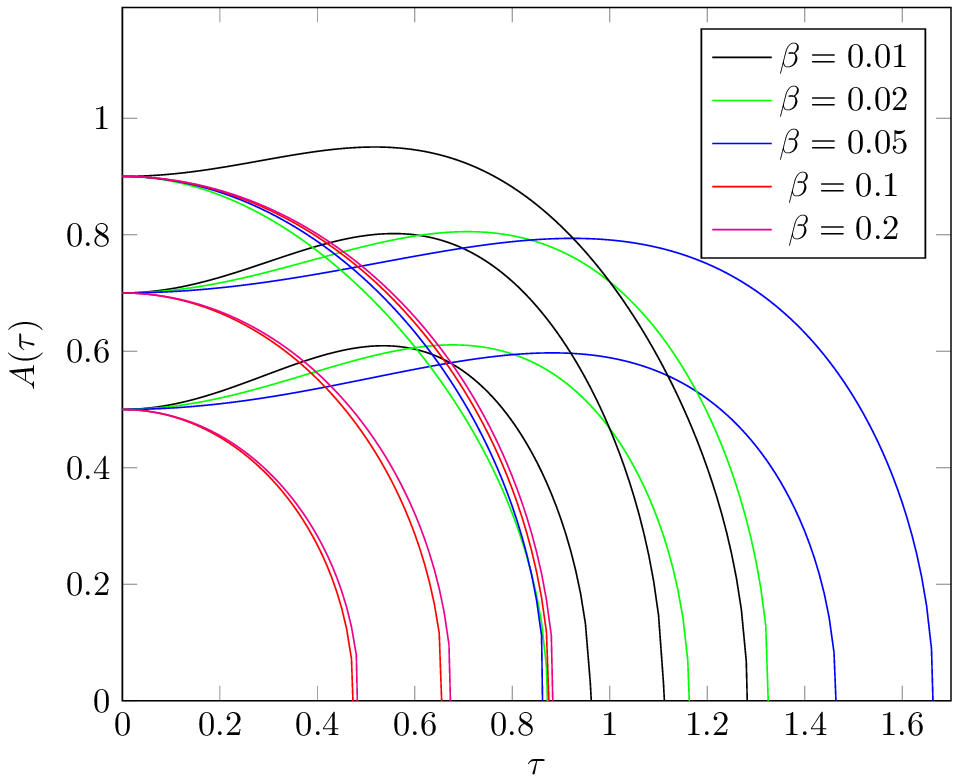}\includegraphics[scale=0.6]{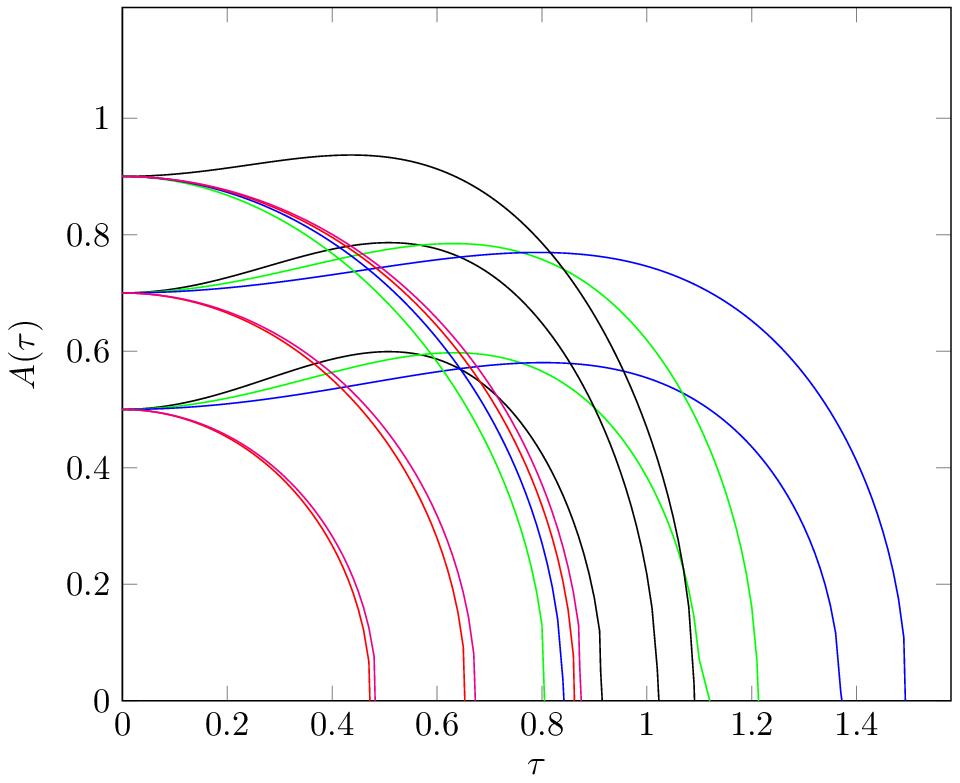}\\
\includegraphics[scale=0.6]{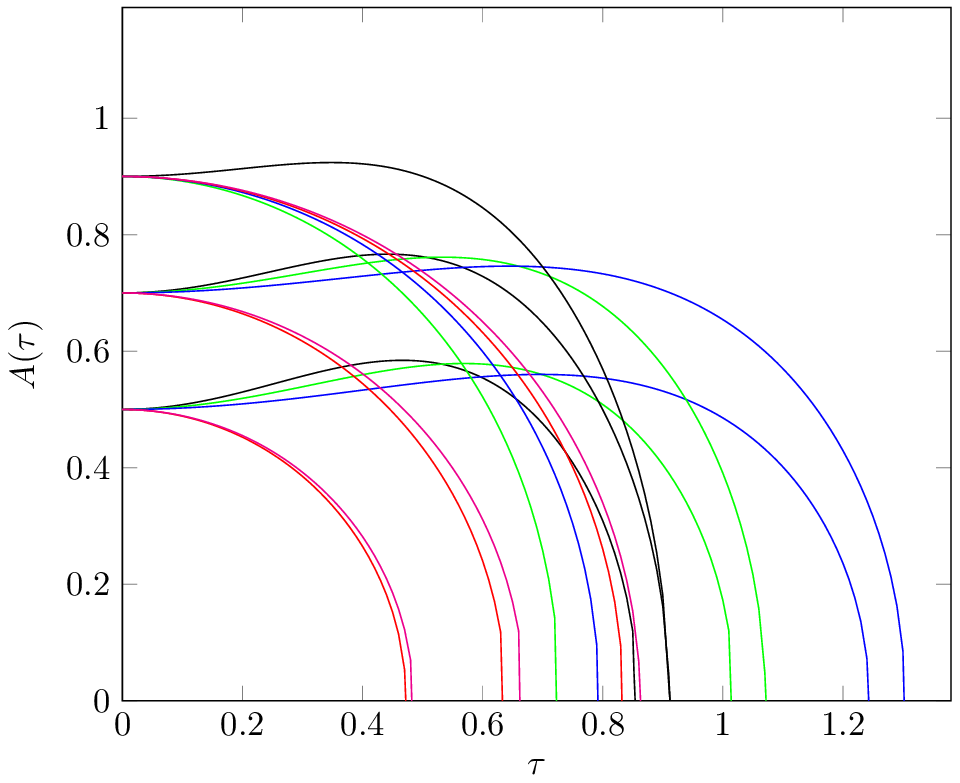}
\end{center}
\caption{Dynamics of scale factor $A(\tau)$ for dust matter (upper
left panel), radiation (upper right panel) and stiff fluid (down
panel) in frames of $R^2$ gravity for various $\beta$ and initial
conditions $H(0)=0$, $\dot{R}(0)=0$, $\rho(0)=1$. Three initial
values for $A(0)$ ($0.5$, $0.7$, $0.9$) are considered.}
\label{sf3}
\end{figure}

We also considered another class of initial conditions. Assuming
$H(0)=0$, $\dot{R}=0$, $\rho(0)=1$ one can vary $A(0)$. Additional
degree of freedom (curvature $R(0)$) allows to satisfy condition
$H(0)=0$. The results are presented on Fig. 4. One can see the
following features. For small values of $\beta$ $A(t)$ increases
and then decreases. The time of collapse grows with $\beta$ in
narrow interval. But for any value of $A(0)$ there is a minimal
value of parameter $\beta$ at which the process of collapse become
similar to previous case for $\beta>\sim 0.1$. Time of collapse
asymptotically tends to $\sim A(0)$ (for previous case we have
$A(0)=1$ and therefore $\tau_{f}=1$).

\section{Conclusion}

We compared the collapse dynamics of homogeneous perfect fluid in
General Relativity and in $R^2$ gravity and formulated the
equations governing various types of fluids in this category,
namely stiff fluid, radiation, dust and the cosmological constant.
The general features are the same for General Relativity and its
above simple modification. As expected stiff fluid collapses to a
singularity faster than the radiation fluid, which in turn, is
faster than the dust fluid collapse. Vacuum energy reduces the
rate of collapse.

We also investigated the collapse of a perfect fluid in the
presence of phantom fluid, one of the candidate of the dark
energy. We choose case $\frac{dV(\phi)}{d\phi}=(V_0+\alpha
t^m)\phi^n$ for potential of scalar field in order to make the
coefficient of $\phi^n$ to play a significant role in determining
of the astrophysical scenario. We obtained the Klein-Gordon
equation which we compared with a general differential equation of
the form Eq.(\ref{anharmonic}) and used a theorem of integrability
condition for the anharmonic oscillator to get an equation which
governs the collapse dynamics of the perfect fluid in phantom
field.

Comparing the gravitational collapse of ordinary perfect fluid
with that in the presence of phantom fluid we concluded that for
values of $n$ outside the interval $(-4,-3)$, there is no
collapse. This observation could be interpreted as follows. The
fluid cannot to collapse if the dark energy has completely
dominated the process. Except for $ n \in (-3,-1)$ we have no
singularities for finite time (for $ n \in (-3,-1)$ there is a Big
Rip singularity).

{When $n \in (-4,-3)$, the collapse is very different from the
usual collapse in the sense that the scale factor $A(t)$
asymptotically tends to zero and singularity takes place at
infinite time. }

{The role of initial value $V_0$ is such that for greater values
of $V_{0}$ for $n<-4$ and $ n \in(-3,-1) $ the scale factor
increases more slowly. Similarly, for $-4<n<-3$ it decreases more
slowly with growing of $V_{0}$.}

In frames of  $f(R)=R+\beta R^2$ theory of gravity gravitational
collapse has some features in comparison with General Relativity.
Firstly due to additional degree of freedom (value of scalar
curvature at the initial moment) one can consider various initial
conditions for collapse. For time derivatives of scalar curvature
and scale factor we take zero values. General feature is that the
collapsing time decreases with increasing of parameter $\beta$ in
narrow interval for equation of state parameter $w<1/3$ and
increases for $w>1/3$. But for $\beta>\beta_{0}$ ($\beta_{0}$
depends from density of collapsing volume) the dynamics of
collapse doesn't depend significantly from $w$. Addition of vacuum
energy leads to increasing of collapsing time as expected.
Depending on value of vacuum energy fraction we have various
dynamics: simple contraction, collapse including several
consecutive stages of expansion and final contraction or expansion
(backward collapse). Non-zero initial curvature leads to dynamics
similar in a case of $R(0)=0$ but for small values of parameter
$\beta$ initial small expansion takes place and then contraction.

There is number of candidates of dark energy which tries to fix
the gap between the observation and theory. However, not all are
expected to be correct and most of them will be discarded sooner
or later. Even in the case of phantom fluid, there are multiple
candidates (different functions $P(X,V)$ in addition to different
potential functions $V(\phi)$) each of them can change the
collapse scenario. {The same is true for different candidates of
modified gravity, like the $f(R)$ gravity}.  Of course for
detailed description of realistic collapse for example in a case
of massive stars or central areas of galaxies one needs to know
the equation of state for collapsing matter. Investigating the
dynamics of gravitational collapse could possibly help us by
shedding some light on the viability of the dark energy  {as well
as modified gravity models} so that some of them could be
discarded, thereby reducing the set of viable models.

\section*{Acknowledgments}

One of authors (KM) wishes to acknowledge the financial support
(Junior Research Fellowship) provided by Council of Scientific and
Industrial Research  (CSIR, India) File
No.:09/919(0031)/2017-EMR-1 for carrying out this research work.
The work of Artyom V. Astashenok is supported by project
1.4539.2017/8.9 (MES, Russia). This research is also supported in
part by MINECO (Spain) FIS2016-76363-P (SDO).

\end{document}